\documentclass[aps,pra,twocolumn,groupedaddress,floatfix,showpacs]{revtex4-1}
\usepackage{amssymb,amsmath}
\usepackage{graphicx}
\usepackage{subfigure}
\usepackage[english]{babel}
\usepackage{float}
\usepackage{color}

\newcommand{\be}{\begin{equation}}
\newcommand{\ee}{\end{equation}}

\begin{document}

\title{Flat bands and $\mathcal{PT}$-symmetry in quasi-one-dimensional lattices}

\author{Mario I. Molina}

\address{Department of Physics, MSI-Nucleus on Advanced Optics, and Center for Optics and Photonics (CEFOP), Faculty of sciences, University of Chile, Santiago, Chile}


\begin{abstract}
We examine the effect of adding $\mathcal{PT}$-symmetric gain and loss terms to quasi 1D lattices (ribbons) that possess flatbands.  We focus on three representative cases: (a) The Lieb ribbon, (b) The kagome ribbon, and (c) The stub Ribbon. In general we find that the effect on the flatband depends strongly on the geometrical details of the lattice being examined. One interesting and novel result that emerge from an analytical calculation of the band structure of the Lieb ribbon including gain and loss,  is that its flatband survives the addition of PT-symmetry for any amount of gain and loss, while for the other two lattices, any presence of gain and loss  destroys the flatbands. For all three ribbons, there are finite stability windows whose size decreases with the strength of the gain and loss  parameter. For the Lieb and kagome cases, the size of this window converges to a finite value. The existence of finite stability windows, plus the constancy of the Lieb flatband are in marked contrast to the behavior of a pure one-dimensional lattice. 
\end{abstract}

\maketitle

\section{INTRODUCTION}

The concept of $\mathcal{PT}$-symmetry has gained considerable attention in recent years. It was started with the	seminal	work of	Bender	and	coworkers \cite{Bender1, Bender2}, who demonstrated that 
non-hermitian Hamiltonians are capable of displaying a purely real eigenvalue spectrum when the system was invariant with respect to the combined operations of parity ($\mathcal{P}$) and time-reversal ($\mathcal{T}$ ) symmetry. When applied to one-dimensional systems, the $\mathcal{PT}$requirement that the imaginary part of the potential term in the Hamiltonian be an odd function, while its real part be even. In a $\mathcal{PT}$-symmetric system, the effects of loss and gain can balance each other and, as a result, give rise to a bounded dynamics. The system thus described can experience a spontaneous symmetry breaking from a $\mathcal{PT}$-symmetric phase (all eigenvalues real) to a broken phase (at least two complex eigenvalues), as the imaginary part of the potential is increased. 

In the case of optics, the paraxial wave equation is formally identical to a Schr\"{o}dinger equation and, as a consequence, the potential is proportional to the index of refraction. In this context, the $\mathcal{PT}$-symmetry requirements translates into the condition that the real part of the refractive index be an even function, while the imaginary part be an odd function in space. 

To date, numerous $\mathcal{PT}$-symmetric systems have been explored in several fields, from optics\cite{Makris1,Makris2,Makris3,Guo,Ruter,Regensburger}, electronic circuits\cite{Schindler}, solid-state and atomic physics\cite{Hatano, Joglekar}, to magnetic metamaterials\cite{Lazarides}, among others. The $\mathcal{PT}$-symmetry-breaking phenomenon has also been observed in several experiments \cite{Guo, Szameit,Ruter,Miroshnichenko}.
It has been sown that a 1D simple periodic lattice with homogeneous couplings and endowed with gain and loss obeying $\mathcal{PT}$-symmetry,  is always in the broken phase of this symmetry and does not have a stable parameter window\cite{Tsironis}. For finite $\mathcal{PT}$-symmetrical lattices with homogeneous couplings, it has been shown that $\mathcal{PT}$-symmetry is preserved inside a parameter window whose size shrinks with the number of lattice sites\cite{Molina}. If one breaks the homogeneity of the couplings, and consider an infinite binary lattice, it was shown that there is a well-defined parameter window where $\mathcal{PT}$-symmetry is preserved\cite{Dmitriev}.

On the other hand, hermitian systems that exhibit flat bands have attracted considerable interest in the past few years, including optical\cite{Apaja, Hyrkas} and photonic lattices\cite{Rechtsman, Vicencio, Thompson}, graphene\cite{Kane, Guinea}, superconductors\cite{Simon,Deng,Deng2, Imada}, 
fractional quantum Hall systems\cite{Tang, Neupert, Yang} and exciton-polariton condensates\cite{Jacqim, Baboux}. The presence of a flat band in the spectrum of a hermitian lattice implies the existence of entirely degenerate states, whose superposition displays no dynamical evolution. This allows the formation of compacton-like structures, that are completely localized in space, constituting a new form of localized state in the continuum. Such states have been recently observed experimentally in an optical waveguide array forming a Lieb lattice in the transversal direction\cite{Vicencio, Thompson}. This raises the possibility that a judicious superposition of these compacton-like states can be used to generate a whole set of diffraction-free modes that can carry information for long distances in an optical waveguide array. It becomes interesting then, to examine the robustness of these localized modes under the presence of balanced loss and gain, obeying 
$\mathcal{PT}$-symmetry. The simplest lattice that is not strictly one-dimensional,  where one can have $\mathcal{PT}$-symmetry, is a quasi one-dimensional one with homogeneous couplings, i.e., a ribbon\cite{PLA}.

In this work we study analytically and numerically, the spectrum and localization properties of three quasi one-dimensional lattices with flat bands (Lieb, kagome and stub), and how their spectra is affected by the presence of gain and loss terms that are $\mathcal{PT}$-symmetric. As we will see, the effect depends strongly on the particulars of the topology of the ribbon being studied. While in the case of the stub and kagome ribbons the presence of gain and loss destroys the flat bands, in the case of the Lieb ribbon, we show analytically that its flat band remains unaltered no matter how large the strength of the gain and loss terms. 

\begin{figure}[t]
\includegraphics[scale=0.75]{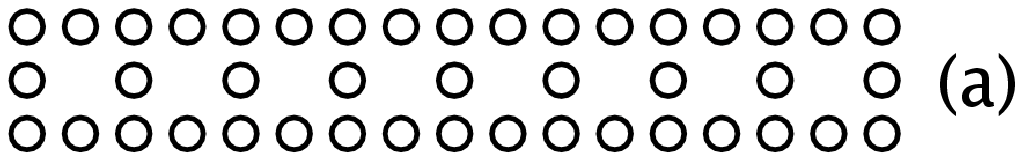}\\
\vspace{0.4cm}
\includegraphics[scale=0.6]{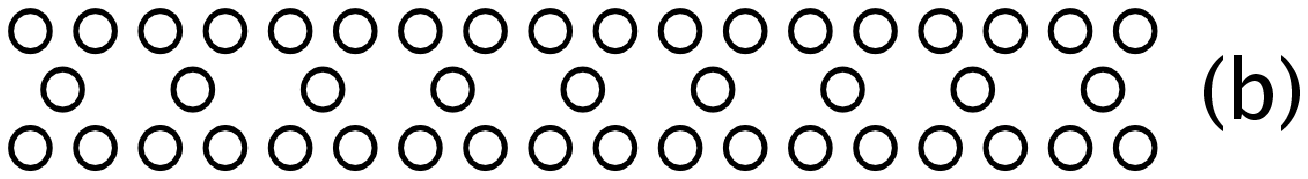}\\
\vspace{0.4cm}
\includegraphics[scale=0.75]{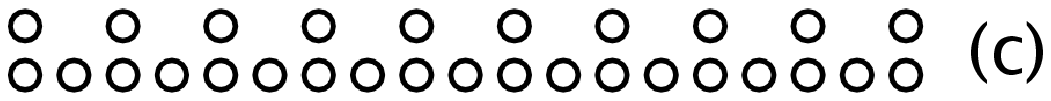}
\caption{lieb(a), kagome(b) and stub(c) ribbons, with homogeneous coupling and in the absence of $\mathcal{PT}$-symmetry. They have infinite extension along the horizontal direction}
\label{fig1}
\end{figure}
\section{The model}
Let us consider a quasi one-dimensional lattice (Ribbon)  representing, for example, a cross section of an optical waveguide array (see Fig. 1). In this context and in the coupled-modes framework, the evolution of the electric field on guide ${\bf n}$ is given by\cite{dnls}
\be
i {d\over{d z}}C_{\bf n} + i \rho_{\bf n} + V \sum_{\bf m} C_{\bf m} = 0\label{eq:1}
\ee
where $C_{\bf n}$ is proportional to the amplitude of the electric field at site ${\bf n}$, $z$ is the propagation coordinate, $\rho_{\bf n}$ is the gain and loss coefficient on site ${\bf n}$, $V$ is the coupling among waveguides, and the sum in Eq.(\ref{eq:1}) is restricted to nearest-neighbors only. Stationary modes are obtained from the {\em ansatz} $C_{\bf n}(z) = C_{\bf n} \exp(i \lambda z)$, where the $C_{\bf n}$ amplitudes obey
\be
-\lambda C_{\bf n} + i \rho_{\bf n} + V \sum_{\bf m} C_{\bf m} = 0\label{eq:2}
\ee
where $\lambda$ is the propagation constant of the mode.  
Fig.1 shows three examples of such ribbons that will be considered in this work. The presence of $\rho_{\bf n}$ leads, in general, to an exponential increase or decrease of the amplitude $C_{\bf n}$ as the mode evolves in ``time'' $z$. However, as was mentioned in the Introduction, there are special cases where the gain and loss terms can be balanced so that the dynamics remain bounded. Such is the case of a $\mathcal{PT}$-symmetric configuration where the value of the $\rho_{\bf n}$ is an odd function in space. The gain and loss term of the three ribbons shown in Fig.1 can be set up as to obey this condition. These ribbons also possess flat bands in their spectrum. What we want to know is the effect of $\mathcal{PT}$-symmetry on those flat bands. To accomplish this , we will examine the spectra of these ribbons as well as the average participation ratio of the states, $PR=(\sum_{n} |C_{n}|^2)^2/\sum_{n} |C_{n}|^4$,  which provides a measure of localization. For a completely localized state, $PR=1$, while for a completely delocalized state $PR=N$.

\begin{figure}[t]
\includegraphics[scale=0.55]{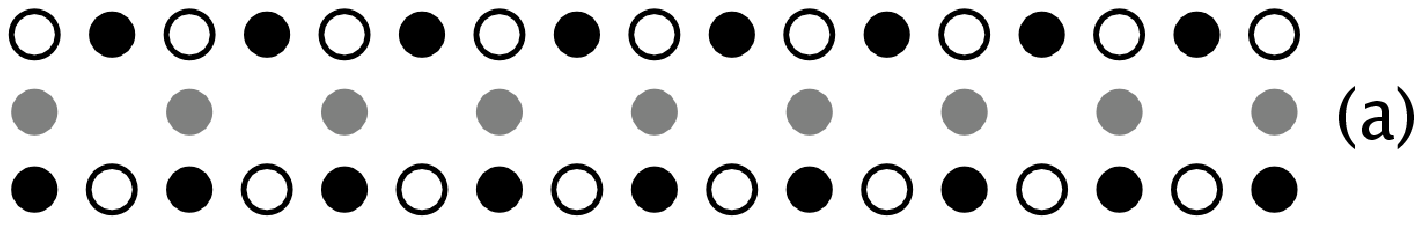}\\
\vspace{.4cm}
\includegraphics[scale=0.53]{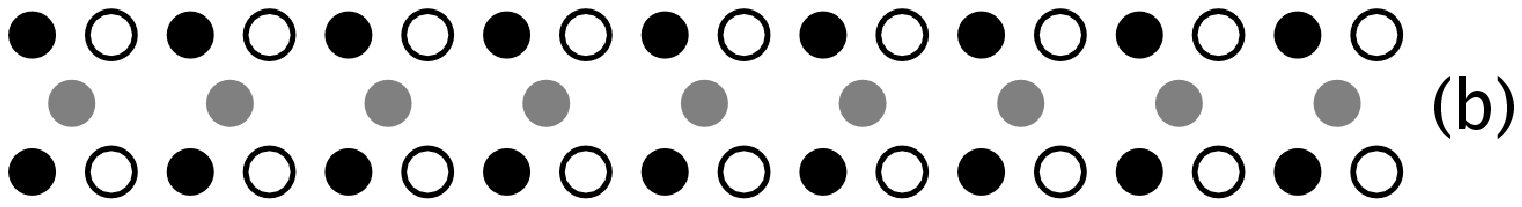}\\
\vspace{0.4cm}
\includegraphics[scale=0.5]{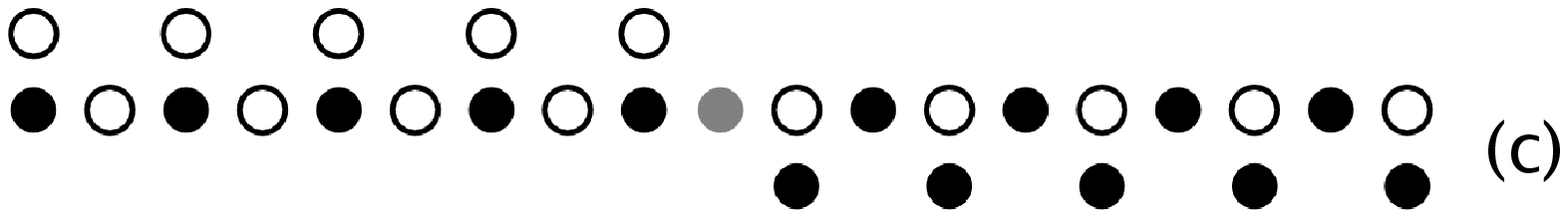}\\
\vspace{0.4cm}
\includegraphics[scale=0.53]{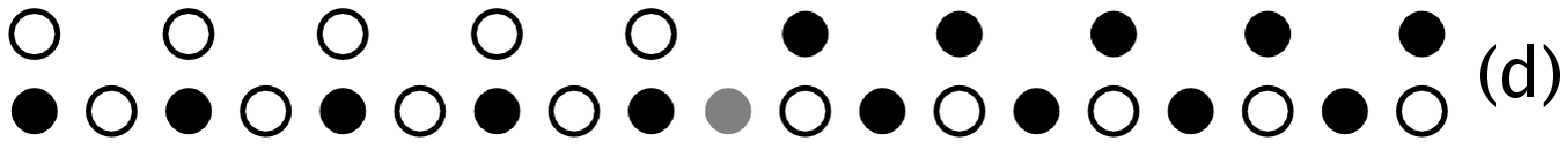}\\

\caption{Ribbons with $\mathcal{PT}$-symmetry: (a) $\mathcal{PT}$-symmetric Lieb ribbon, (b) $\mathcal{PT}$-symmetric  kagome ribbon, (c) $\mathcal{PT}$-symmetric stub ribbon and (d) topologically equivalent $\mathcal{PT}$-symmetric stub ribbon. Black(white) circles denote loss (gain), while the gray circles represent absence of gain and loss.}
\label{fig2}
\end{figure}
\vspace{0.5cm}

\subsection{The Lieb ribbon}
\label{lieb}

The Lieb ribbon is shown in Fig.1 (top row). It consists essentially of a depleted square lattice ribbon. Its unitary cell contains five units. In the absence of gain and loss, one obtains five bands
\begin{eqnarray}
\lambda &=& 0\nonumber\\
\lambda &=& \pm \sqrt{2 (1+\cos(k))} V\nonumber\\
\lambda &=& \pm \sqrt{4 + 2 \cos(k)} V.
\end{eqnarray}
Thus, out of the five bands, we have the flat band $\lambda=0$. The modes belonging to this band have zero group velocity, which leads to a sharp transverse localization. These compacton-like modes are able to propagate along the guide without diffraction. The reason for this localization is a geometric phase cancellation among nearby sites. 
Some examples of such modes are can be found in ref.\cite{dany}.

Let us now incorporate $\mathcal{PT}$-symmetric gain and loss into the system. There many ways to achieve this, and we take the simplest one, depicted on the upper row of Fig.2. For this configuration, the five coupled equations incorporating $\mathcal{PT}$-symmetry 
\begin{figure}[t]
\includegraphics[scale=0.75]{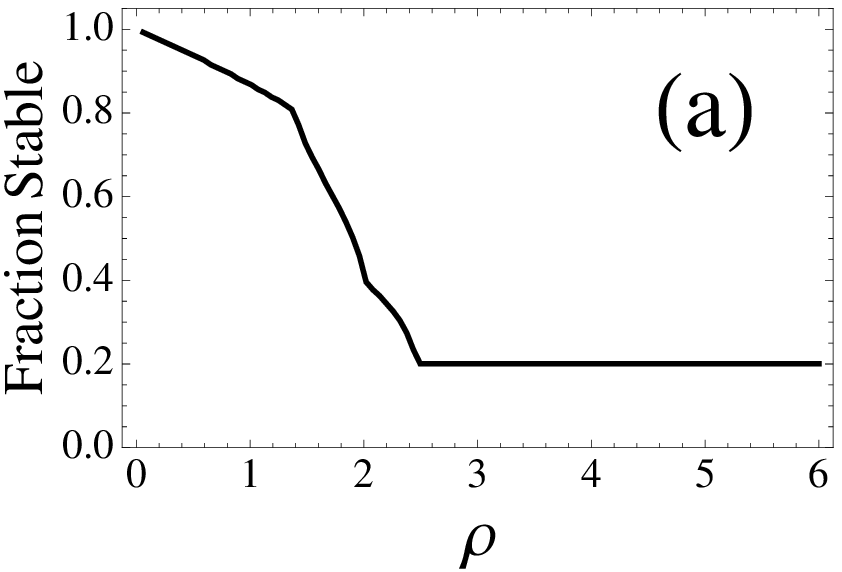}\\
\vspace{.4cm}
\includegraphics[scale=0.75]{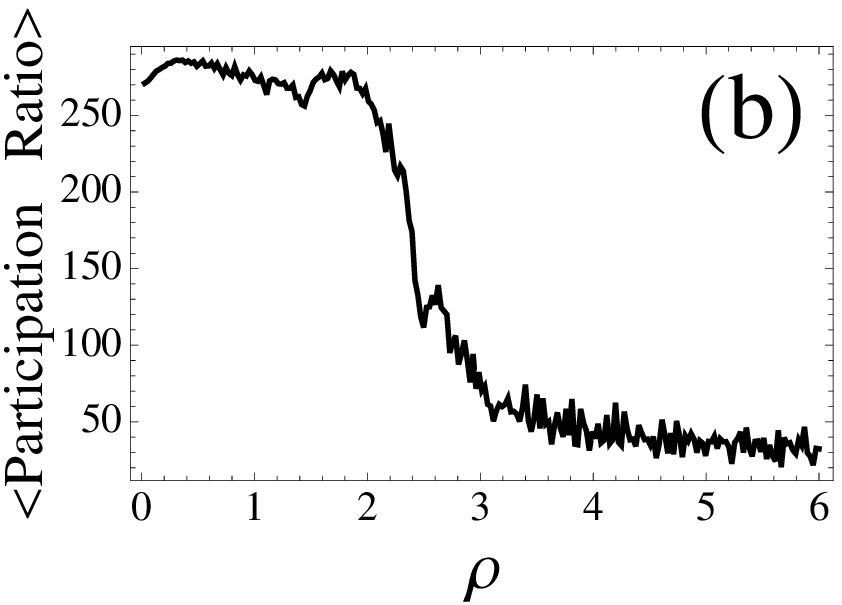}\\
\vspace{0.4cm}
\caption{Lieb ribbon: (a) Fraction of stable modes Im[$\lambda]=0$ as a function of the gain and loss parameter. (b) Average of the participation ration $\langle PR\rangle$ over all stable states, as a function of the gain and loss parameter.} 
\label{fig3}
\end{figure}
lead to the five complex bands
\begin{eqnarray}
\lambda &=& 0\nonumber\\
\lambda &=& \pm \sqrt{2 (1+\cos(k))V^2-\rho^2}\ \nonumber\\
\lambda &=& \pm \sqrt{(4 + 2 \cos(k)) V^2-\rho^2}\ .
\end{eqnarray}

As we can see, the flat band $\lambda=0$ still remains and is,  therefore, unaltered by the presence of $\mathcal{PT}$-symmetric gain and loss terms. The other four (dispersive) bands are real at small $\rho$ values, but become purely imaginary at critical values of $\rho$ (at $\rho=2$ and $\rho=\sqrt{6}$). If one now concentrates on the fraction of stable states (i.e., Im[$\lambda]=0$), that is, the fraction with purely real eigenvalues, we can predict that, as the gain and loss parameter $\rho$ is increased, this fraction will decrease as well and, at large $\rho$ values will approach a constant value stemming from the flat band states (which are also stable, of course). They constitute a $1/5$ of the total number of states. Thus the stable fraction should approach asymptotically a value of $0.2$. Figure 3 shows the stable fraction and the average (over stable states) of the participation ratio $\langle PR\rangle$, as a function of the gain and loss parameter. While at small values of $\rho$ the $\langle PR\rangle$ stays nearly constant, it begins to decay rapidly past $\rho\sim \sqrt{6}$ and large oscillations appear. This indicates a tendency towards localization. Since at high $\rho$ values, only the flat band remains as the only stable band, the participation ratio is bounded from below to the value of $4$, which is the simplest flat band eigenstate composed of a four-site ring that is consistent with $\mathcal{PT}$-symmetry\cite{dany}.

\subsection{The kagome ribbon}
\label{kagome}
The kagome ribbon is shown at the middle of Fig.1. It has five sites in its unit cell, which implies five bands. In the absence of gain and loss ($\rho=0$) they are given by
\begin{eqnarray}
\lambda &=& -2\ V\nonumber\\
\lambda &=& \pm \sqrt{2 (1+\cos(k))}\ V\nonumber\\
\lambda &=& ( 1\pm \sqrt{3 + 2 \cos(k)} )\ V.
\end{eqnarray}
Thus, we have the flat band $\lambda=-2 V$. When gain and loss are added, it is no longer possible to extract the bands in closed form as we did for the Lieb lattice. A numerical examination of all eigenvalues reveals that as soon as $\rho$ differs from zero, the flat band is lost. The fraction of stable states, that is, those states with Im[$\lambda]=0$, as well as their participation ratios $\langle PR\rangle$ as a function of the gain and loss parameter $\rho$, are shown in Fig.4. 
\begin{figure}[t]
\includegraphics[scale=0.75]{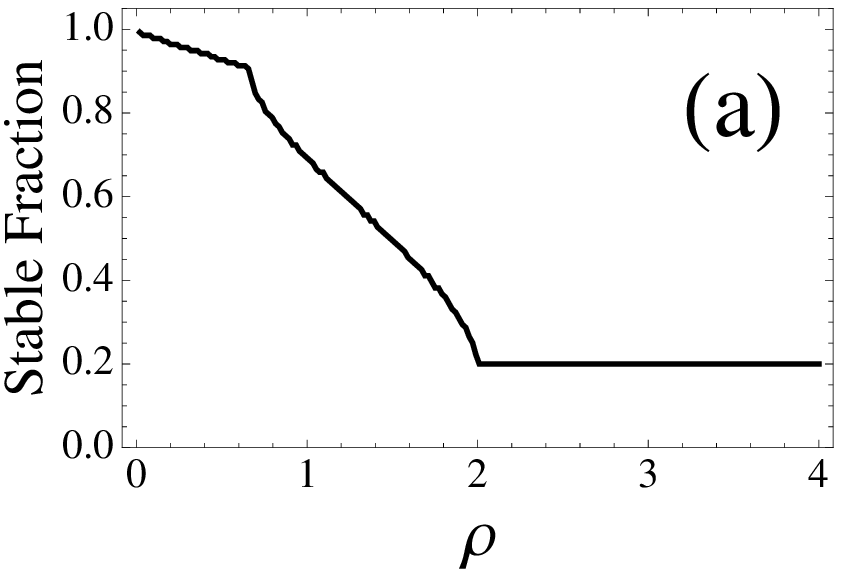}\\
\vspace{.4cm}
\includegraphics[scale=0.75]{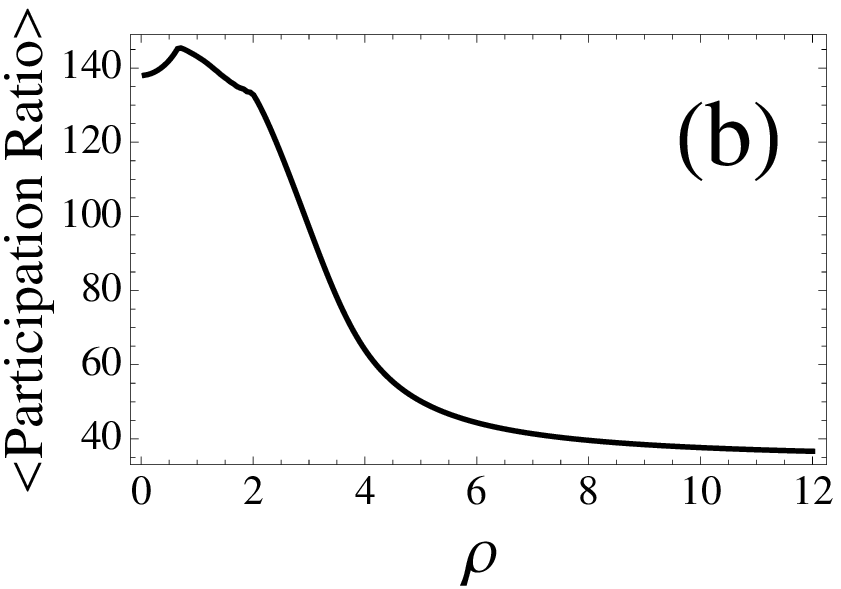}\\
\vspace{0.4cm}
\caption{Kagome ribbon: (a) Fraction of stable modes Im[$\lambda]=0$ as a function of the gain and loss parameter. (b) Average of the participation ration over all stable states, as a function of the gain and loss parameter.} 
\label{fig4}
\end{figure}
The general tendency of Figs. 3 and 4 is the same. In both cases the stable fraction and the participation ratio decrease  with $\rho$. Since in this case we no longer have a flat band, the asymptotic fraction of stable states is only due to the presence of a finite percentage of states with Im[$\lambda]=0$.
\begin{figure}[t]
\includegraphics[scale=0.75]{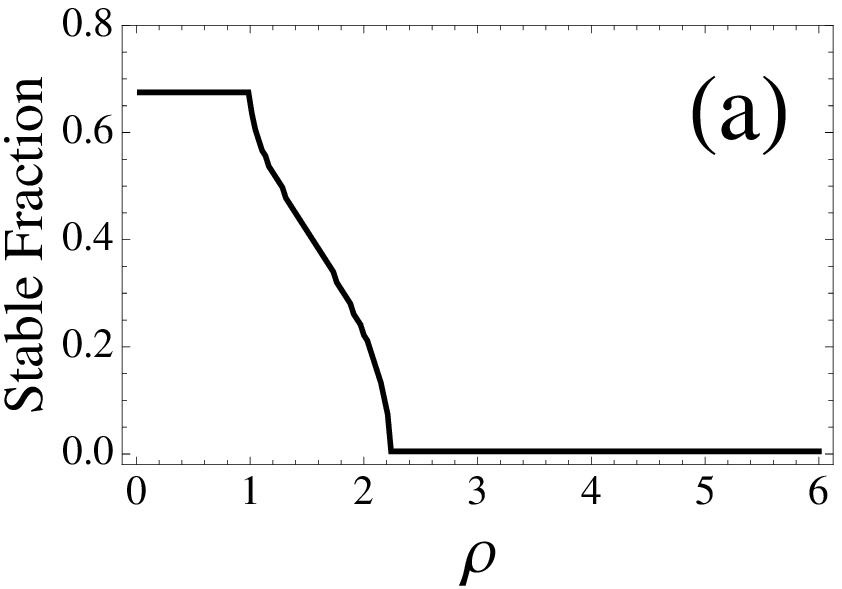}\\
\vspace{.4cm}
\includegraphics[scale=0.75]{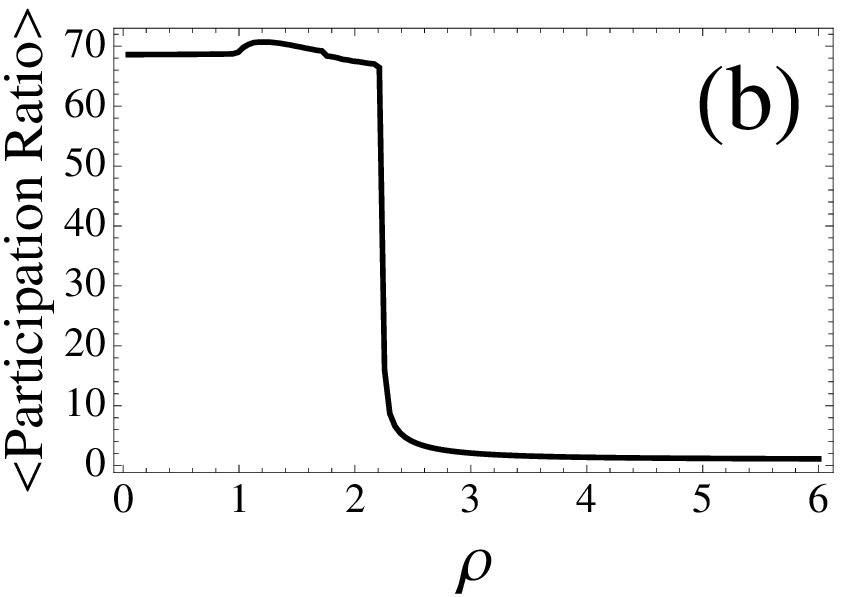}\\
\vspace{0.4cm}
\caption{Stub ribbon: (a) Fraction of stable modes Im[$\lambda]=0$ as a function of the gain and loss parameter. (b) Average of the participation ration over all stable states, as a function of the gain and loss parameter.} 
\label{fig5}
\end{figure}

\subsection{The stub ribbon}
\label{stub}
The stub ribbon is shown at the bottom row of Fig.1. Its unitary cell has three sites. In the absence of gain and loss, this leads to three real bands:
\begin{eqnarray}
\lambda &=& 0\nonumber\\
\lambda &=& \pm \sqrt{3 + 2 \cos(k)}\ V\nonumber\\
\end{eqnarray}
where, as in the Lieb case, we have a flat band $\lambda=0$. A simple $\mathcal{PT}$-symmetric configuration for this lattice is shown at the bottom  part on Fig.2c. It is topologically equivalent to the one shown in Fig. 2d. As we can see, roughly speaking, the lattice has been split into two halves and a closed-form calculation of the eigenvalues from Eq.(\ref{eq:2}) is not possible. Numerical examination of all eigenvalues for varying ribbon lengths reveals that the flat band disappears as soon as $\rho$ is different from zero.

The stable fraction $\langle PR\rangle$ has an interesting behavior: It remains constant until a first critical $\rho$ value is reached. Then it drops with an increase in $\rho$ all the way down to zero, at a second critical value. On the other hand, the participation ratio remains more or less constant with an increase in $\rho$, until reaching the second critical $\rho$ value mentioned before, where it drops abruptly, converging to unity at large gain and loss values. In this case we see a really abrupt transition from relatively extended modes (on average) to highly localized modes.

\section{Discussion}

In this work we have examined the spectral properties of several quasi-one-dimensional lattices (Lieb, kagome and stub) that, in the absence of gain and loss, feature a flat band. We have  incorporated $\mathcal{PT}$-symmetric gain and loss terms and examined the changes in their spectra. The results show that while there are common trends for all of them, there are also features that are present only in each case. Perhaps the most interesting analytical result is that a Lieb ribbon maintains its flat band, regardless of the strength of the gain and loss term. This is quite surprising since usually, the addition of $\mathcal{PT}$-symmetry leads to a stability window which shrinks with the strength of gain and loss. But for the Lieb lattice, the system not only remains stable, but keeps its original flat band for any $\mathcal{PT}$-symmetric gain and loss amount. A common feature for the three cases is the decrease of the stable fraction with an increase in gain and loss. While for the Lieb and kagome ribbon this fraction remain finite at large values of gain and loss, for the stub lattice it vanishes at certain $\rho$ value.
The  average participation value of all ribbons also decreases with an increase in gain and loss, reaching a finite value at hight $\rho$ values. For the stub lattice in particular, the $\langle PR\rangle$ approaches unity. Now this $\langle PR\rangle$ is a rough estimate and only measures the general tendency towards localization. As the gain and loss parameter $\rho$ increases, the stationary wavefunction  seems to concentrate more and more power ($\sum_{n} |C_{n}|^2$) at certain sites causing the decrease in $\langle PR \rangle$. As long as this power concentration is finite, the system will be dynamically stable.

We conclude that the spectral properties of a given ribbon depends on its geometry, and that the addition of $\mathcal{PT}$-symmetry to a ribbon possessing a flat band
will result in most cases in a complete destruction of the flat band. An exception to this behavior is the Lieb ribbon where its flat band shows a remarkable robustness to $\mathcal{PT}$-symmetry. This feature isolates the Lieb ribbon as a possible candidate for a stable long-distance image transmission system. Its quasi-one-dimensional geometry, makes its fabrication  possible by means of the laser-written waveguide technique\cite{szameit}. 

\section{Acknowledgements}

This work was partially supported by FONDECYT grant 1120123, Programa ICM P10-030-F and Programa de Financiamiento Basal de CONICYT (FB0824/2008).

\end{document}